\documentclass[12pt]{article}

\usepackage{epsfig}

\begin{document}
\renewcommand{\thefootnote}{\fnsymbol{footnote}}



\def\vare{\varepsilon}
\def\bz{\bar{z}}
\def\bw{\bar{w}}


\def\CA{{\cal A}}
\def\CC{{\cal C}}
\def\CD{{\cal D}}
\def\CE{{\cal E}}
\def\CF{{\cal F}}
\def\CG{{\cal G}}
\def\CI{{\cal I}}
\def\CT{{\cal T}}
\def\CM{{\cal M}}
\def\CN{{\cal N}}
\def\CP{{\cal P}}
\def\CL{{\cal L}}
\def\CV{{\cal V}}
\def\CS{{\cal S}}
\def\CW{{\cal W}}
\def\CY{{\cal Y}}
\def\CS{{\cal S}}
\def\CO{{\cal O}}
\def\CP{{\cal P}}
\def\CN{{\cal N}}


\def\IC{\mathbb{C}}
\def\ID{\mathbb{D}}
\def\IH{\mathbb{H}}
\def\IP{\mathbb{P}}
\def\IR{\mathbb{R}}
\def\IZ{\mathbb{Z}}


\def\a{\alpha}
\def\b{\beta}
\def\g{\gamma}
\def\d{\delta}
\def\e{\epsilon}
\def\ve{\varepsilon}
\def\i{\iota}
\def\k{\kappa}
\def\l{\lambda}
\def\m{\mu}
\def\n{\nu}
\def\th{\theta}
\def\vth{\vartheta}
\def\r{\rho}
\def\vr{\varrho}
\def\s{\sigma}
\def\vs{\varsigma}
\def\t{\tau}
\def\u{\upsilon}
\def\w{\omega}
\def\z{\zeta}
\def\G{\Gamma}



\def\half{\frac{1}{2}}
\def\imp{\Longrightarrow}
\def\dint#1#2{\int\limits_{#1}^{#2}}
\def\goto{\rightarrow}
\def\para{\parallel}
\def\brac#1{\langle #1 \rangle}
\def\del{\nabla}
\def\grad{\nabla}
\def\curl{\nabla\times}
\def\div{\nabla\cdot}
\def\p{\partial}


\def\Tr{{\rm Tr}}
\def\det{{\rm det}}
\def\im{{\rm Im~}}
\def\re{{\rm Re~}}


\begin{titlepage}
\hfill\parbox{4cm}{ KIAS-P02016 \\ hep-th/0204055}
\vspace{15mm}
\baselineskip 8mm

\begin{center}
{\LARGE \bf On Decoupling of Massless Modes \\ in NCOS
Theories}
\end{center}

\baselineskip 6mm
\vspace{10mm}
\begin{center}
Seungjoon Hyun\footnote{\tt hyun@phya.yonsei.ac.kr}$^a$,
Sangmin Lee\footnote{\tt sangmin@newton.skku.ac.kr}$^{b,c}$ and
Hyeonjoon Shin\footnote{\tt hshin@kias.re.kr}$^{b,c}$
\\[5mm]
$^a${\sl Institute of Physics and Applied Physics, Yonsei University,
Seoul 120-749, Korea}
\\
$^b${\sl BK 21 Physics Research Division and Institute of Basic Science}
\\
{\sl Sungkyunkwan University, Suwon 440-746, Korea}
\\
$^c${\sl School of Physics, Korea Institute for Advanced Study, Seoul
130-012, Korea}
\end{center}

\thispagestyle{empty}


\vskip 3cm

\begin{center}
{\bf Abstract}
\end{center}
\noindent

We revisit the decoupling phenomenon of massless modes in 
the noncommutative open string (NCOS) theories. 
We check the decoupling by explicit computation in (2+1) 
or higher dimensional NCOS theories
and recapitulate the validity of the decoupling 
to all orders in perturbation theory.

\vspace{15mm}
\end{titlepage}

\baselineskip 6.5mm
\renewcommand{\thefootnote}{\arabic{footnote}}
\setcounter{footnote}{0}


\section{Introduction}

Noncommutative open string (NCOS) theories are 
defined by a scaling limit 
in which one takes an electric field on a brane to its critical value
and at the same time scale $g_s$ and $\a'$ appropriately 
so that the closed strings decouple and the open strings 
experience maximal noncommutativity in the electric direction 
\cite{ncos1, ncos2, ncos3, GMSS}.
Their properties including S-duality and T-duality 
have been studied extensively \cite{kle}-\cite{hs}. 

It is of great interest to compare the dynamics of NCOS theories and their 
S-duals, but our limited understanding of the strong 
coupling dynamics allows only a few explicit checks. 
One such example is the decoupling phenomenon of massless modes in NCOS 
theories \cite{kle}. The authors of \cite{kle} observed that
 the S-dual super-Yang-Mills theories, 
in (1+1) or (3+1) dimensions for example,
contain massless $U(1)$ fields which are free and decouple 
from the rest of the theory when their momenta have components 
only in the electric directions. 
S-duality implies then that 
the massless modes of NCOS theories also should be free. 
They showed that this is indeed the case 
by computing some amplitudes explicitly and also by 
giving a general argument based on holomorphy of world-sheet correlators. 
Subsequently, Ref. \cite{her} gave more details of decoupling 
of the (1+1) dimensional NCOS and Ref. \cite{oogo} 
generalized the holomorphy argument to prove that vanishing 
of the relevant amplitudes is exact to all order in perturbation theory.
 
So far, explicit computations have been limited to (1+1) dimensional NCOS. 
In the present work, we compute the amplitudes in (2+1) 
or higher dimensional NCOS theories and check that 
the decoupling phenomenon remains to be valid in higher dimensions.  
We also review the proof of the decoupling to all order given in \cite{oogo} 
from a slightly different point of view.


\section{Explicit Computation}

Higher dimensional NCOS theories are more complicated than the 
(1+1)-dimensional one in two ways. First, the string modes can carry momentum
in the non-electric directions. Second, the massless spectrum contains 
not only the transverse scalars but also gauge bosons. 
The goal of this section is to confirm by explicit computation that 
the decoupling phenomenon holds for arbitrary  world-volume dimensions.

As a warm-up exercise, we first compute the amplitudes in the bosonic theory. 
We will see that they share all the essential features of decoupling 
with the superstring theory.

\subsection{Bosonic String}

Following \cite{kle, her}, we check the amplitudes involving massless modes 
and tachyons. The vertex operators are given by
\begin{eqnarray}
V_T &=& G_o e^{i p \cdot X},
            \nonumber \\
V_\phi &=& \frac{G_o}{\sqrt{\alpha'_e}}
    \zeta_i \partial X^i e^{ip \cdot X},
            \nonumber \\
V_A &=& \frac{G_o}{\sqrt{\alpha'_e}} e_\alpha \partial X^\alpha
    e^{ip \cdot X},
\end{eqnarray}
where the scalars represent the transverse fluctuation of the
world-volume ($\zeta_i$) and the gauge bosons are polarized along
the world-volume ($e_\alpha$). The world-sheet correlators are
\begin{eqnarray}
\langle X^\alpha (y) X^\beta (y') \rangle
  &=& - 2 \alpha'_e \eta^{\alpha \beta} \ln | y-y'|
      + \frac{i}{2} \theta \epsilon^{\alpha \beta} {\rm sign} (y-y'),
                \nonumber \\
\langle \partial X^i (z) \partial X^j (z') \rangle
  &=& -\frac{\alpha'_e}{2} \frac{\eta^{ij}}{(z-z')^2},
\end{eqnarray}
where $\a = 0, \cdots, p$ and $i = p+1, \cdots, 25$. 
$\epsilon^{01} = - \epsilon^{10} = 1, \epsilon^{\a\b} = 0$ otherwise and 
$\theta = 2 \pi \alpha'_e E$. 
In the NCOS limit, $E \rightarrow 1$ and hence $\theta \rightarrow 2
\pi \alpha'_e$.
We introduce the dimensionless Mandelstam variables,
\begin{eqnarray}
s &=& - \alpha'_e (p_1+p_2)^2,  \nonumber
\\
t &=& - \alpha'_e (p_1+p_4)^2,  \nonumber
\\
u &=& - \alpha'_e (p_1+p_3)^2.
\end{eqnarray}
Noncommutativity gives rise to phase factors. 
To each of the three inequivalent cyclic ordering, we associate a
phase factor in the following way 
($p\times q = \theta \epsilon^{\a\b} p_\a q_\b$),
\begin{eqnarray}
\gamma_1 &=& \cos \left[ \frac{1}{2} ( p_1 \times p_4 +p_2 \times p_3 )
          \right],
                    \nonumber \\
\gamma_2 &=& \cos \left[ \frac{1}{2} ( p_1 \times p_2 +p_3 \times p_4 )
          \right],
                    \nonumber \\
\gamma_3 &=& \cos \left[ \frac{1}{2} ( p_1 \times p_3 -p_2 \times p_4 )
      \right].
\end{eqnarray}

The simplest amplitude of our interest is the scattering of a scalar 
off a tachyon. For obvious reasons, the result is formally 
the same as the (1+1)-dimensional answer given in \cite{kle,her}. 
The transverse scalars are taken to have momenta $p_1$ and
$p_4$ and polarization tensors $\zeta$ an $\zeta'$, respectively, 
while tachyons have $p_2$ and $p_3$.  
Then the four-point scattering amplitude 
${\cal A}_{\phi\phi TT}$ is given by
\begin{equation}
\label{ttss}
{\cal A}_{\phi\phi TT}
 = -\frac{G_o^2}{\alpha'_e} \zeta \cdot \zeta'
    \left[ B(-1-t,-u) \gamma_1
          + B(-1-t,-s) \gamma_2
          + B(-u,-s) \gamma_3
    \right]~,
\end{equation}
where $B(a,b)$ is the beta function and Mandelstam variables
satisfy $s+t+u=-2$. It is convenient to rewrite this amplitude
by using an identity for the gamma function, $
\Gamma(z) \Gamma(1-z) = \pi / \sin ( \pi z )~.  $ The right hand side
of Eq.~(\ref{ttss}) then becomes
\begin{equation}
\label{ttss2}
 - \frac{G_o^2}{\alpha'_e} \zeta \cdot \zeta'
    \frac{B(-1-t,-s)}{ \sin (\pi u) } \Phi_-~.
\end{equation}
Here we have defined $\Phi_\pm$ as
\begin{equation}
\Phi_\pm \equiv \sin (\pi s) \gamma_1 + \sin (\pi u) \gamma_2
      \pm \sin (\pi t) \gamma_3~.
\end{equation}
The factors $\Phi_\pm$, which combine the $\gamma$ factors due to the
space-time noncommutativity and sine functions of Mandelstam
variables, are crucial for showing the decoupling of massless modes
from the rest of the NCOS modes. 
Other amplitudes involving massless modes 
also turn out to contain the factors $\Phi_\pm$. 
Since higher dimensional NCOS theories have dynamical gauge 
fields in addition to the transverse scalars, 
we still need to compute $\CA_{\phi\phi\phi\phi}$, $\CA_{AA\phi\phi}$, 
$\CA_{AATT}$, $\CA_{AAAA}$. 
The standard recipe for computing the amplitudes gives
\begin{enumerate}

\item
Four scalars with momenta $p_n$ and 
polarizations $\zeta_n$ ($n=1,...,4$) : 
\begin{eqnarray}
{\cal A}_{\phi \phi \phi \phi}
 &=& \frac{G_o^2}{2 \alpha'_e}
        \left[ \zeta_1 \cdot \zeta_2 \zeta_3 \cdot \zeta_4
                        \frac{ B(1-t,1-u) }{ \sin (\pi s)}
              + \zeta_1 \cdot \zeta_3 \zeta_2 \cdot \zeta_4
                        \frac{ B(1-t,1-s) }{ \sin (\pi u)}
        \right.
                                        \nonumber \\
 & &   \hspace{7mm}
        \left. + \zeta_1 \cdot \zeta_4 \zeta_2 \cdot \zeta_3
                        \frac{ B(1-s,1-u) }{ \sin (\pi t)}
        \right] \Phi_+ ~
\end{eqnarray}

\item
Two gauge bosons with $(p_1, e )$ and $(p_4, e')$,  
and two transverse scalars ($p_2, \zeta)$ and $(p_3, \zeta')$ :
\begin{eqnarray}
{\cal A}_{AA \phi \phi}
 &=& \frac{G_o^2}{2\alpha'_e}
    \frac{\zeta \cdot \zeta'}{\sin (\pi t)}
                \nonumber \\
 &\times &
   \bigg\{
    \bigg[  e \cdot e'
      -2 \alpha'_e ( e \cdot p_2 e' \cdot p_2
            +e \cdot p_3 e' \cdot p_3)
        \bigg] B(1-s,1-u) 
                \nonumber \\
 & & +2 \alpha'_e  \Big[
            e \cdot p_2 e' \cdot p_3
                   B (2-u,-s)
                + e \cdot p_3 e' \cdot p_2
               B (2-s,-u)
                  \Big] 
   \bigg\} \Phi_+
\end{eqnarray}

\item
Two gauge bosons with $(p_1, e)$ and $(p_4, e')$
and two tachyons :
\begin{eqnarray}
{\cal A}_{AATT}
 &=& \frac{G_o^2}{\alpha'_e}
        \frac{1}{\sin (\pi t)}
                                \nonumber \\
 &\times & \bigg\{
2 \alpha'_e  \Big[
                        e \cdot p_2 e' \cdot p_3
                           B (1-u,-1-s)
                        + e \cdot p_3 e' \cdot p_2
                           B (1-s,-1-u)
                  \Big]         
                                \nonumber \\
 &&  
+ \bigg[ e \cdot e'
          -2 \alpha'_e ( e \cdot p_2 e' \cdot p_2
                        +e \cdot p_3 e' \cdot p_3)
        \bigg] B(-s,-u) 
     \bigg\} \Phi_-
\end{eqnarray}

\end{enumerate}
The expression for $\CA_{AAAA}$ is more complicated. 
For our purposes, it suffices to note that it is proportional to 
$\Phi_+$.
We observe that the appearance of the factors $\Phi_\pm$ is universal 
in the sense that the scattering amplitudes between massless states 
among themselves are always proportional to $\Phi_+$ and 
the ones between massless modes and massive modes are proportional 
to $\Phi_-$. 
In order to check the vanishing of the amplitude
in the NCOS limit, it is sufficient to show that these factors vanish.


We show the vanishing of $\Phi_-$ first. There are three types of physical 
processes to consider: 
(a) Pair annihilation of massless particles and 
a subsequent pair creation of massive ones 
(b) Forward scattering of a massless particle off a massive one 
(c) Backward scattering.
Our choice of center of mass momenta and the resulting phase factors 
are summarized in the following table. 

\begin{center}
\begin{tabular}{lccc}
& Pair annihilation & Forward scattering & Backward scattering \\
$p_1$ & $(p,p,\vec{0})$ & $(p,p,\vec{0})$ & $(p,p,\vec{0})$ \\
$p_4$ & $(p,- p,\vec{0})$ & $-(p,p,\vec{0})$ & $-(p,-p,\vec{0})$ \\
$p_2$ & $-(p, q ,\vec{k})$ & $(e,-p,\vec{k})$ & $(e,-p,\vec{0})$ \\
$p_3$ & $-(p,-q,-\vec{k})$ & $-(e,-p,\vec{k})$ & $-(e,p,\vec{0})$ \\
$\g_1$ & $\cos(\pi E (u+1))$ & 1 & $\cos(\pi E (u+1))$ \\
$\g_2$ & $\cos(\pi E (s+1))$ & 1 & $\cos(\pi E (s+1))$ \\
$\g_3$ & 1 & $\cos(\pi E (s+1))$ & 1
\end{tabular}
\end{center}

\noindent
For instance, in the case of pair annihilation, 
\begin{equation}
\Phi_- = \sin(\pi s) \cos \left( \pi E (u+1) \right)
	+\sin(\pi u) \cos \left( \pi E (s+1) \right)
        -\sin(\pi t), 
\end{equation}
which vanishes in the NCOS limit ($E \rightarrow 1$) 
due to the fact that $s+t+u = -2$.
The backward scattering is shown to vanish in the same way. 
Note that the forward scattering vanishes even before taking the NCOS limit.

The vanishing of $\Phi_+$ can be verified similarly. The only change 
from the previous case is that the mass-shell condition is different. 
If we replace $(u+1)$ by $u$ and $(s+1)$ by $s$ in the phase factors 
$\gamma_{1,2,3}$, 
the above table becomes valid for the case at hand. 
This replacement together with the condition that $s+t+u=0$ for massless modes
make sure that $\Phi_+$ also vanishes.

\subsection{Superstring}

Strictly speaking, the NCOS theories and their S-duals are well-defined 
only in Type II superstring theories.
However, as we saw in the previous subsection, 
the reason for the decoupling is mainly kinematical and
is not sensitive to supersymmetry. 
Therefore, we expect that the main feature of the amplitude we found in 
the bosonic case will persist in the superstring case. 

The massless spectrum is basically the same as before. 
Instead of the tachyon of the bosonic theory, we consider
the first massive states. 
Our notation for the polarization tensors is as follows.
\begin{itemize}
\item $e_\alpha$ : massless gauge field
\item $\zeta_i$  : massless transverse scalar or $SO(9-p)$ vector
\item $\epsilon_{\alpha\beta}$ : massive $m^2 = 1/\alpha'_e$ 
$SO(1,p)$ tensor
\item $\xi_{ij}$ : massive $m^2=1/\alpha'_e$ $SO(9-p)$ tensor
\end{itemize}
\noindent
The vertex operator for the gauge field on the brane with polarization 
$e_\alpha$ is
\begin{eqnarray}
V_{-1} &=& G_o e^{-\phi} e \cdot \psi e^{ik \cdot X}(y)~,
    \nonumber
\\
V_0 &=& \frac{G_o}{\sqrt{\alpha'_e}} e_\alpha
    (\partial X^\alpha + i \alpha'_e k \cdot \psi \psi^\alpha)
    e^{ik \cdot X} (y).
\end{eqnarray}
The vertex operator for the transverse massless scalar 
is defined in the same way with $e_\a$ replaced by $\zeta_i$. 
The vertex operator for the massive tensor $N_{ij}$ with mass square
$1/\alpha'_e$ and polarization transverse to the brane, $\xi_{ij}$ is
\begin{eqnarray}
V_{-1} &=& \frac{G_o}{\sqrt{\alpha'_e}} e^{-\phi}
        \xi_{ij} \psi^i \partial X^j e^{ik \cdot X} (y)~,
        \nonumber
\\
V_0 &=& \frac{G_o}{\alpha'_e} \xi_{ij}
    ( \partial X^i \partial X^j -\alpha'_e \psi^i \partial \psi^j
     + i \alpha'_e (k \cdot \psi) \psi^i \partial X^j),
     e^{ik \cdot X} (y)
\end{eqnarray}
where $k^2= -1/\alpha'_e$.
The vertex operator for the massive tensor $M_{\alpha\beta}$ 
is defined similarly.
The world-sheet propagator for the fermion is
\begin{equation}
\langle \psi^\mu (z) \psi^\nu (z') \rangle = \frac{\eta^{\mu\nu}}{z-z'} .
\end{equation}
A straightforward calculation following the standard recipe gives the 
amplitudes. There are seven different amplitudes involving massless states. 
Here we present five of them. 
\begin{enumerate} 

\item
Four transverse scalars with momenta and polarizations ($p_n$, $\zeta_n$) 
($n=1, \cdots, 4$):
\begin{eqnarray}
{\cal A}_{\phi\phi\phi\phi}
 &=& - \frac{G_o^2}{\alpha'_e} \frac{\Gamma(-t)\Gamma(-s)}{\Gamma(1-t-s)}
      \frac{1}{\sin (\pi u)} 
					\nonumber \\
 &\times& \Big( su \zeta_1 \cdot \zeta_4 \zeta_2 \cdot \zeta_3
	       +ut \zeta_1 \cdot \zeta_2 \zeta_3 \cdot \zeta_4
	       +st \zeta_1 \cdot \zeta_3 \zeta_2 \cdot \zeta_4
    	  \Big) \Phi_+
\end{eqnarray}

\item
Gauge fields with ($p_1$, $e$) and ($p_4$, $e'$)
and transverse scalars with ($p_2$, $\zeta$) and
($p_3$, $\zeta'$):
\begin{equation}
{\cal A}_{AA \phi \phi}
 = - \frac{G_o^2}{\alpha'_e} \frac{\Gamma(-t)\Gamma(-s)}{\Gamma(1-t-s)}
     \frac{1}{\sin (\pi u)} \zeta \cdot \zeta' 
    \Big( su e \cdot e' - 2 \alpha'_e u e \cdot p_2 e' \cdot p_3
         - 2 \alpha'_e s e \cdot p_3 e' \cdot p_2 
    \Big) \Phi_+
\end{equation}

\item
Transverse scalars with ($p_1$, $\zeta$) and
($p_4$, $\zeta'$) and massive $SO(9-p)$ tensors with
($p_2$, $\xi$) and ($p_3$, $\xi'$):
\begin{eqnarray}
{\cal A}_{\phi\phi N N }
 &=& \frac{G_o^2}{2\alpha'_e} \frac{\Gamma(-t) \Gamma(-s)}{\Gamma(3-t-s)}
     \frac{1}{\sin (\pi u)}
    \Big[ - us (1-s)(1-u) \zeta \cdot \zeta' {\rm Tr} (\xi^T \cdot \xi') 
					\nonumber \\
 & & + tu (1-u) \zeta \cdot \xi^T \cdot \xi' \cdot \zeta'
     + ts (1-s) \zeta' \cdot \xi^T \cdot \xi' \cdot \zeta
					\nonumber \\
 & & + stu (1-s) \zeta \cdot \xi' \cdot \xi^T \cdot \zeta'
     + stu (1-u) \zeta' \cdot \xi' \cdot \xi^T \cdot \zeta 
     \Big] \Phi_-
\end{eqnarray}

\item
Transverse scalars ($p_1$, $\zeta$) and
($p_4$, $\zeta'$) and massive $SO(1,p)$ tensor with 
($p_2$, $\epsilon$) and ($p_3$, $\epsilon'$):
\begin{eqnarray}
{\cal A}_{\phi \phi MM}
 &=& - \frac{G_o^2}{2 \alpha'_e} \frac{\Gamma(-t) \Gamma(-s)}{\Gamma(3-t-s)}
     \frac{1}{\sin (\pi u)} \zeta \cdot \zeta'
   \Big[ su (1-s) (1-u) {\rm Tr}(\epsilon' \cdot \epsilon^T)
					\nonumber \\
 & & + 2 \alpha'_e su(1-u) p_4 \cdot \epsilon' \cdot \epsilon^T \cdot p_1
     + 2 \alpha'_e su(1-s) p_1 \cdot \epsilon' \cdot \epsilon^T \cdot p_4 
   \Big] \Phi_-
\end{eqnarray}

\item
Gauge fields with ($p_1$, $e$) and ($p_4$, $e'$) and 
massive $SO(9-p)$ tensor with ($p_2$, $\xi$) and ($p_3$, $\xi'$): 
\begin{eqnarray}
{\cal A}_{AANN}
  &=& -\frac{G_o^2}{2 \alpha'_e} 
      \frac{\Gamma(-t)\Gamma(-s)}{\Gamma(3-t-s)}
      \frac{1}{\sin (\pi u)} {\rm Tr} (\xi^T \cdot \xi')
    \Big[
        su (1-s) (1-u) e \cdot e'
					\nonumber \\
  & &	 + 2 \alpha'_e su (1-s) e \cdot p_2 e' \cdot p_3
	 + 2 \alpha'_e su (1-u) e \cdot p_3 e' \cdot p_2
    \Big] \Phi_-
\end{eqnarray}

\end{enumerate}
The expressions for the other two amplitudes, namely,  
$\CA_{AAAA}$ and $\CA_{AAMM}$ are more complicated, 
but for our purposes it suffices to note that  $\CA_{AAAA} \propto \Phi_+$ and 
$\CA_{AAMM} \propto \Phi_-$.
We realize that the factors $\Phi_\pm$ appear 
in the same way as in the bosonic case. 
Thus it is clear that the decoupling is valid 
in all NCOS theories.

\vskip 10mm

\section{Vanishing of the Loop Amplitudes}

A general argument for decoupling was given in \cite{kle} 
based on holomorphy of the world-sheet correlators in the NCOS limit.
The authors of \cite{oogo} generalized this argument to all orders 
using a first order formalism to describe NCOS. 
The formalism made it possible to prove the decoupling 
without computing higher loop amplitudes explicitly.
In this section, we present a version of the proof to all orders 
with more emphasis on the explicit form of the world-sheet correlators.

\subsection{Disk}

When the background electric field is turned on in the $x^1$ direction, 
the world sheet propagator on a disk becomes
\begin{eqnarray}
\label{tree1}
\langle X^\a(z,\bz) X^\b(w,\bw) \rangle 
&=& 
- \frac{\a'_e}{2} \eta^{\a\b} 
\left[ (1-E^2) \ln|z-w|^2 + (1+E^2) \ln|z-\bw|^2 \right]
\nonumber \\
&&
-\frac{\a'_e}{2} \e^{\a\b} (2E) \ln\frac{z-\bw}{\bz-w}.
\end{eqnarray}
We obtain the open-string propagator 
by taking the insertion points to the boundary.
Suppose we take $w$ to the boundary first. In the light cone coordinates 
$X^\pm = X^0 \pm X^1$, the propagator becomes
\begin{equation}
\langle X^\pm (z,\bz) X^\mp(0) \rangle 
= 2 \a'_e (1 \pm E) \ln(z)  + 2 \a'_e (1 \mp E) \ln(\bz).
\end{equation}
It follows that in the NCOS limit ($E\rightarrow 1$), 
$\langle X^+(z) X^-(0)\rangle$ becomes a holomorphic function of $z$ and 
$\langle X^-(z) X^+(0)\rangle$ anti-holomorphic.
This is not quite a coincidence. 
When we act Laplacian on the propagator (\ref{tree1}), 
the $\ln |z-w|^2$ term produces a delta function. 
In the NCOS limit, this term drops out due to the $(1-E^2)$ factor, 
so the propagator satisfies a source-free Laplace equation. 
The most general solution to the source-free equation 
is a sum of a holomorphic function and an anti-holomorphic function. 
Now the boundary condition (after Wick rotation)
\begin{equation}
- \partial_n X^0 + i E \partial_t X^1 = 0 , \;\;\;
\partial_n X^1 - i E \partial_t X^0 = 0
\end{equation}
implies that 
$\partial_{\bz} \langle X^+(z,\bz) X^-(0)\rangle = \partial_{z} \langle X^-(z,\bz) X^+(0)\rangle =0$. This boundary condition can be satisfied only if 
either the anti-holomorphic or the holomorphic part vanishes.

\begin{figure}[!h]
\centerline{\hbox{\psfig{figure=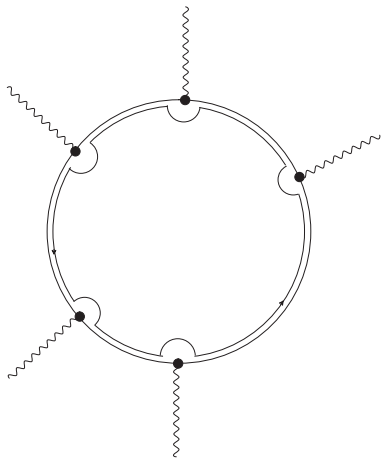,height=60mm}}}
\caption{{\footnotesize The massless vertex operator is integrated while 
others are kept fixed.} 
{\footnotesize Wavy lines represent the branch-cuts emanating from 
the other vertices.}} \label{d2f1}
\end{figure}

When a massless mode has momentum only in the electric direction, 
the on-shell condition implies $p_+ p_- = 0$. Assume that $p_-=0$ so 
that the anti-holomorphic $\langle X^- X^+ \rangle$ drops out.  
If we make the branch cuts lie outside the disk, the amplitude becomes an 
integral of a holomorphic function. Then by shrinking the contour, 
we can show that the amplitude vanishes.

\subsection{Annulus}

The world-sheet propagator on an annulus in the presence of 
background magnetic field is given in \cite{dorn}-\cite{chau}. 
Repeating the same exercise with electric field, 
taking one of the insertion points to a boundary 
and taking the NCOS limit, we find that 
\begin{equation}
\label{annulus1}
\langle X^+(z,\bz) X^-(0) \rangle 
= 4 \a'_e \left[ \ln\left(\frac{\theta_1(z|iT)}{\theta'_1(0|iT)}\right)
+ \frac{\pi}{T} (z^2 +z) \right].
\end{equation}
It is holomorphic as expected from   
the argument given in the previous section. 
The quadratic term and the linear term in $z$ in (\ref{annulus1}) ensure
periodicity for $z \rightarrow z+iT$.  
In fact, the propagator does not have to be strictly periodic 
because a constant shift drops out due to momentum conservation 
when we actually compute the amplitude.

\begin{figure}[!h]
\centerline{\hbox{\psfig{figure=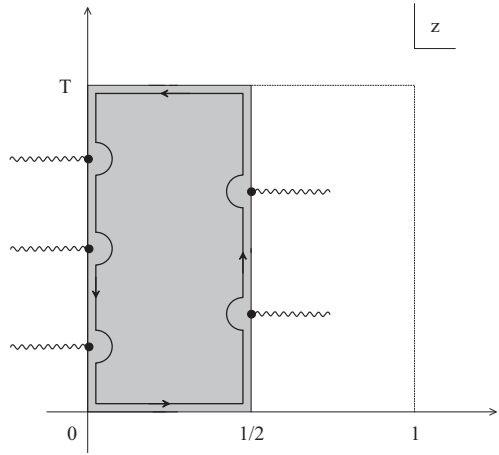,height=60mm}}}
\caption{{\footnotesize The world-sheet boundaries are $x=0, 1/2$. 
The $y=0$ and $y=T$ lines are identified. }} \label{d2f2}
\end{figure}

As before, we can put all the branch cuts outside the annulus. 
Since the integration measure is obviously the same for the two boundaries, 
we can again deform the contour and show that the amplitude vanishes.

\subsection{Higher Loops}

Generalization to higher loop is straightforward. 
Ref. \cite{klp} gives some details of higher loop propagator 
in the presence of noncommutativity. 
We refer the reader to \cite{klp} for our notation. 
The propagator in the NCOS limit is
\begin{equation}
\label{higher1}
\langle X^+(z,\bz) X^-(c=\bar{c}) \rangle 
= 4 \a'_e \left[ \ln(E(z,c|iT))+ \pi (T^{-1})^{ij}\Omega_i \Omega_j \right],  
\end{equation}
where $E(z,c)$ is the prime form, $\Omega_i (z,c) = \int_c^z \omega_i$ and 
$\omega_i$ is an Abelian differential. 
In the one-loop case, the prime form reduces to the theta function 
and the abelian differential to $dz$, hence we are back to the annulus answer 
(up to an irrelevant linear term). 

One may wonder whether the integration measure for the massless vertex 
operator is the same for all boundaries thereby allowing deformation of 
the contour. The behavior of the prime form under coordinate transformation 
and modular transformation dictates that the integration measure is 
simply proportional to $dz$ restricted to boundary. In particular, 
the integration measure does not depend on the moduli parameters. 
As such, all the boundaries are on the same footing and 
the integration measure should thus be the same.

\section*{Acknowledgements}

We are grateful to Youngjai Kiem for helpful discussions.
The work of S.H. was supported in part by grant No. 2000-1-11200-001-3
from the Basic Research Program of the Korea Science and
Engineering Foundation.


\end{document}